\documentclass[12pt,thmsa]{article}
\usepackage{amssymb}

\usepackage{sw20lart}

\input tcilatex
\begin{document}

\title{Fractional Kinetics for Relaxation and Superdiffusion in Magnetic Field}
\author{A. V. Chechkin$^{a}$, V. Yu. Gonchar$^{a}$, M. Szyd\l owski$^{b}$ \\
$^{a}$Institute for Theoretical Physics\\
National Science Center\\
''Kharkov Institute of Physics and Technology''\\
Akademicheskaya st. 1, 61108, Kharkov, Ukraine,\\
$^{b}$Astronomical Observatory,\\
Jagiellonian University,\\
30-244, Orla 171,\\
Cracow, Poland}
\maketitle

\begin{abstract}
We propose fractional Fokker-Planck equation for the kinetic description of
relaxation and superdiffusion processes in constant magnetic and random
electric fields. We assume that the random electric field acting on a test
charged particle is isotropic and possesses non-Gaussian Levy stable
statistics. These assumptions provide us with a straightforward possibility
to consider formation of anomalous stationary states and superdiffusion
processes, both properties are inherent to strongly non-equilibrium plasmas
of solar systems and thermonuclear devices. We solve fractional kinetic
equations, study the properties of the solution, and compare analytical
results with those of numerical simulation based on the solution of the
Langevin equations with the noise source having Levy stable probability
density. We found, in particular, that the stationary states are essentially
non-Maxwellian ones and, at the diffusion stage of relaxation, the
characteristic displacement of a particle grows superdiffusively with time
and is inversely proportional to the magnetic field.

\end{abstract}

\newpage

\newpage

\section{\strut Introduction}

\strut Anomalous random motions and related transport phenomena are
ubiquitous in nature. In these phenomena the laws of normal diffusion
(ordinary Brownian motion) are altered, e.g., the mean square no longer
increases linearly in time, but instead grows slower (sub-diffusion) or
faster (super-diffusion) than the linear function. There are a lot of
examples from very different fields of applications, see reviews \cite
{Bouchaud}, \cite{Metzler} and references therein. The anomalous random
motions often exhibit long-time and/or space dependence as well as
multi-scaling, or multifractal, behavior \cite{Mandelbrot}, \cite{Feder}.
These circumstances require to go beyond the theory of (relatively simple)
random Markovian processes as well as beyond the theory of (mono)fractal, or
self-affine, processes. The systems, in which anomalous random motions occur
are usually essentially non-linear and, in this sense, the random motions
are non-linear ones; this circumstance again greatly complicates the problem
of an adequate statistical description.

The two basic anomalous fractal random motions are of particular importance,
namely, fractional Brownian motion \cite{van Ness}, and the Levy motion,
whose theory has begun from the works of P.\ Levy \cite{Levy}. The former
motion is characterized by long-range time correlations, whereas the latter
one is characterized by non-Gaussian statistics ; in this case the
increments of the process may be independent (Levy stable processes or
ordinary Levy motion \cite{Skorokhod}) or have an infinite span of
interdependence (fractional Levy motion) \cite{Maejima}, \cite{Samorodnitsky}%
, \cite{Chechkin1} .

The theory of Levy stable distributions and stable processes naturally
serves as the basis for probabilistic description of the Levy motion, since
stable distributions obey Generalized Central Limit Theorem, thus
generalizing Gaussian distribution \cite{Gnedenko}. However, the application
of the theory of stable processes is limited because of the infiniteness of
the mean square and discontinuity of the sample paths. The finite sample
size and boundary effects play a decisive role, thus modifying stable
probability laws (''truncated Levy distributions'') \cite{Stanley}, and
violating the property of self-affinity (''spurious multi-affinity'')\cite
{Chechkin2}.

The peculiarity of anomalous random motions is that they cannot be described
by the standard Fokker-Planck equation because the basic assumptions,
namely, the Markov property and the local homogeneity of the space do not
hold in these cases. The use of differential equations with partial
fractional derivatives is a perspective way for describing such processes.
One of the implementations of such an approach is the use of different forms
of fractional kinetic Fokker-Planck equation or the corresponding forms of
the Master equation with fractional derivatives. Recently, kinetic equations
with fractional partial derivatives have attracted attention as a possible
tool for the description of diffusion and relaxation phenomena, see review 
\cite{Metzler} and references therein. However, fractional calculus is far
from being a regular tool in physics community, and the solutions to
fractional kinetic equations are known in a very few cases. The development
of the theory requires, from one hand, the development of microscopic
foundations of fractional kinetics and, from the other hand, the development
of powerful regular methods for solutions to fractional equations.

\strut The various processes in space and thermonuclear plasmas could serve
as important applications of fractional kinetics. Indeed, many of the
current challenges in solar system plasmas as well as in plasmas of
thermonuclear devices arise from the fundamentally multiscale and nonlinear
nature of plasma fluctuation and wave processes. Anomalous diffusion and
plasma heating, particle acceleration and macroscopic transfer processes
require to go beyond the ''traditional'' plasma kinetic theory. Fractional
kinetics can be useful for describing such processes, just as it occurs in
other fields of applications. Our paper is a step in this direction. We
consider the motion of a charged Levy particle in a constant external
magnetic field and random electric field obeying non-Gaussian Levy
statistics. Our problem is a natural generalization of the classical example
of the motion of a charged Brownian particle \cite{Korsunoglu}. We solve the
fractional Fokker-Planck equation with fractional velocity derivative, study
the relaxation processes in phase and real spaces as well, and estimate
fractional moments of energy and coordinate. We also perform numerical
modelling based on the numerical solution to the Langevin equations and
demonstrate qualitative agreement between analytical and numerical results.

\section{\strut Fractional Fokker-Planck equation for charged particle in
magnetic field}

The history of fractional Fokker-Planck equation (FFPE) for the probability
density function (PDF) $f(\vec{r},\vec{v},t)$ in the phase space goes back
to the papers by West and Seshadri \cite{West}, and by Peseckis \cite
{Peseckis}. Here we recall briefly the arguments used when deriving FFPE. It
is well known that usually the derivation of classical kinetic equations for
the Brownian motion is based on the assumption of the finiteness of the
second moments of the PDF. This way is not useful here, because, as we shall
see, the second moments diverge. Thus, it is expedient to explore the method
used by Chandrasekhar \cite{Chandrasekhar} for the derivation of the
Fokker-Planck equation for the Brownian motion. His method does not require
the finiteness of the second moments. In fact, for fractional kinetics, the
modification of Chandrasekhar's method was proposed for the first time in
Ref. \cite{Peseckis}. Here we proceed mainly along the derivation from Ref. 
\cite{Chechkin3}. We consider a test charged particle with the mass $m$ and
the charge $e$, embedded in constant external magnetic field $\vec{B}$ and
subjected to stochastic electric field $\mathcal{\vec{E}}\left( t\right) .$%
We also assume, as in the classical problem for the charged Brownian
particle \cite{Korsunoglu}, that the particle is influenced by the linear
friction force $-\nu m\vec{v}$, $\nu $ is the friction coefficient. For this
particle the Langevin equations of motion are 
\begin{eqnarray}
\frac{d\vec{r}}{dt} &=&\vec{v}\,\quad ,  \tag{2.1} \\
\frac{d\vec{v}}{dt} &=&\frac{e}{mc}[\vec{v}\times \vec{B}]-\nu \vec{v}+\frac{%
e}{m}\mathcal{\vec{E}}\quad .  \nonumber
\end{eqnarray}

The statistical properties of the field $\mathcal{\vec{E}}\left( t\right) $
are assumed to be as follows.

1. $\mathcal{\vec{E}}\left( t\right) $is homogenous and isotropic.

2. $\mathcal{\vec{E}}\left( t\right) $ is a stationary white Levy noise.

The first assumption is the usual one when dealing with the motion of
charged particle in a random electric field. In subsequent Sections we
consider two possibilities:

(i) $\mathcal{\vec{E}}\left( t\right) $ is a 2-dimensional ($d=2$) isotropic
field in the direction perpendicular to external magnetic field; in this
case the motion along the magnetic field is neglected, and

(ii) $\mathcal{\vec{E}}\left( t\right) $ is a 3-dimensional ($d=3$) field.

The second assumption has a profound meaning.

Indeed, if $\mathcal{\vec{E}}\left( t\right) $ is a white Gaussian noise,
then we encounter with a classical Brownian problem and, by using Eqs.
(2.1), we arrive at the Fokker-Planck equation and get, as the consequences,
Maxwell stationary PDF over the velocity, exponential relaxation to the
Maxwell PDF and the normal diffusion law for the particle motion in the real
space, as well. Instead, the non-Gaussian Levy statistics of the random
force in the Langevin equation (2.1) provides us with a simple and
straightforward, at least, from the methodical viewpoint, possibility to
consider abnormal diffusion and non-Maxwell stationary states, both
properties are inherent to strongly non-equilibrium plasmas of solar system
and thermonuclear devices. Returning to Eq. (2.1), it follows from the
assumption 2 that the process, which is an integral of $\mathcal{\vec{E}}%
\left( t\right) $ over some time lag $\Delta t$,

\begin{equation}
\vec{L}\left( \Delta t\right) =\int\limits_{t}^{t+\Delta t}dt^{\prime }%
\mathcal{\vec{E}}\left( t^{\prime }\right) \text{ ,}  \tag{2.2}
\end{equation}
is an $\alpha $-stable isotropic process with stationary independent
increments \cite{Skorokhod}, \cite{Samorodnitsky}, whose characteristic
function is

\begin{equation}
\hat{\omega}_{L}\left( \vec{k},\Delta t\right) =\exp \left( -D_{\mathcal{E}%
}\left| \vec{k}\right| ^{\alpha }\Delta t\right) \text{ ,}  \tag{2.3}
\end{equation}
where $\alpha $ is called the Levy index, $0<\alpha \leq 2$ , and the
positive parameter $D_{\mathcal{E}\text{ }}$has the physical meaning of the
intensity of the random electric field. If $\alpha =2$ , Eq. (2.3) is a
characteristic function of the Wiener process, and, after applying
Chandrasekhar's procedure we arrive at the Fokker-Planck equation. But if $%
\alpha <2$ then, by applying the procedure described in detail for the
one-dimensional case in Ref. \cite{Chechkin3}, we arrive at the fractional
Fokker-Planck equation for the charged particle in the magnetic field and
random electric field:

\begin{equation}
\frac{\partial f}{\partial t}+\vec{v}\frac{\partial f}{\partial \vec{r}}%
+\Omega \left[ \vec{v}\times \vec{e}_{z}\right] \frac{\partial f}{\partial 
\vec{v}}=\nu \frac{\partial }{\partial \vec{v}}\left( \vec{v}f\right)
-D\left( -\Delta _{\vec{v}}\right) ^{\alpha /2}f\text{ ,}  \tag{2.4}
\end{equation}
where $\Omega =eB/mc$ , $D=e^{\alpha }D_{\mathcal{E}}/m^{\alpha }$ and $%
\left( -\Delta _{\vec{v}}\right) ^{\alpha /2}$ is the fractional Riesz
derivative over the velocity. This operator is defined through its Fourier
transform as

\begin{equation}
\left( -\Delta _{\vec{v}}\right) ^{\alpha /2}f\left( \vec{r},\vec{v}%
,t\right) \div \left| \vec{k}\right| ^{\alpha }\hat{f}\left( \vec{\varkappa},%
\vec{k},t\right)  \tag{2.5}
\end{equation}
where $\hat{f}$ is the characteristic function,

\begin{equation}
\hat{f}\left( \vec{\varkappa},\vec{k},t\right) =\left\langle \exp \left( i%
\vec{\varkappa}\vec{r}+i\vec{k}\vec{v}\right) \right\rangle \text{ ,} 
\tag{2.6}
\end{equation}
the brackets $\left\langle ...\right\rangle $ imply statistical averaging.

An explicit representation of the Riesz derivative is realized through
hypersingular integral, see the monograph \cite{Samko} containing detailed
presentation of the Riesz differentiation. We also note that at $\alpha =2$
Eqs. (2.4), (2.5) are reduced to the Fokker-Planck equation of the Brownian
motion. In the next Sections we get the solution to Eq. (2.4) and consider
physical consequences.

\section{Solution to fractional Fokker-Planck equation}

\strut \strut In this Section we solve Eq. (2.4) with initial condition

\begin{equation}
f\left( \vec{r},\vec{v},t=0\right) =\delta \left( \vec{r}-\vec{r}_{0}\right)
\delta \left( \vec{v}-\vec{v}_{0}\right)  \tag{3.1}
\end{equation}
in the infinite phase space. We pass to the characteristic function (2.6),
which obeys the equation\strut 
\begin{equation}
\frac{\partial \hat{f}}{\partial t}+\left( -\vec{\varkappa}+\Omega \left( 
\vec{k}\times \vec{b}\right) +\nu \vec{k}\right) \frac{\partial \hat{f}}{%
\partial \vec{k}}=-D\left| \vec{k}\right| ^{\alpha }\vec{f}.  \tag{3.2}
\end{equation}
with the initial condition

\begin{equation}
f\left( \vec{\varkappa},\vec{k},t=0\right) =\exp \left( i\vec{\varkappa}\vec{%
r}_{0}+i\vec{k}\vec{v}_{0}\right) .  \tag{3.3}
\end{equation}
Equation (3.2) can be solved by the method of characteristics. The equations
of characteristics are

\begin{equation}
\frac{d\vec{k}}{dt}=\nu \vec{k}+\Omega \left( \vec{k}\times \vec{b}\right) -%
\vec{\varkappa}\,\,\,,  \tag{3.4a}
\end{equation}

\begin{equation}
\frac{d\hat{f}}{dt}=-D\left| \vec{k}\right| ^{\alpha }\vec{f}.  \tag{3.4b}
\end{equation}
Denote $\vec{k}^{\prime }\left( t^{\prime }\right) $ the value of $\vec{k}$
at time instant $t^{\prime }$. Then, the solution to Eq. (3.4a) can be found
after lengthy, but straightforward, calculations:

\[
\vec{K}=e^{\nu \left( t-t^{\prime }\right) }\left\{ \left( \vec{K}^{\prime }%
\vec{b}\right) \vec{b}+\vec{b}\times \left( \vec{K}^{\prime }\times \vec{b}%
\right) \cos \Omega \left( t-t^{\prime }\right) \right. 
\]

\strut 
\begin{equation}
\left. +\left( \vec{K}^{\prime }\times \vec{b}\right) \sin \Omega \left(
t-t^{\prime }\right) \right\}  \tag{3.5a}
\end{equation}

\strut

\[
\vec{K}^{\prime }=e^{\nu \left( t^{\prime }-t\right) }\left\{ \left( \vec{K}%
\vec{b}\right) \vec{b}+\vec{b}\times \left( \vec{K}\times \vec{b}\right)
\cos \Omega \left( t^{\prime }-t\right) \right. 
\]

\strut 
\begin{equation}
\left. +\left( \vec{K}\times \vec{b}\right) \sin \Omega \left( t^{\prime
}-t\right) \right\}  \tag{3.5b}
\end{equation}

\strut

\strut

where $\vec{K}=\vec{k}-\vec{G}$ , $\vec{K}^{\prime }=\vec{k}^{\prime }-\vec{G%
}\,$\thinspace $\,,$

\begin{equation}
\vec{G}=\frac{\left( \vec{\varkappa}\vec{b}\right) }{\nu }+\frac{\nu }{%
\Omega _{1}^{2}}\left( \vec{b}\times \left( \vec{\varkappa}\times \vec{b}%
\right) \right) +\frac{\Omega }{\Omega _{1}^{2}}\left( \vec{b}\times \vec{%
\varkappa}\right) \,\,,  \tag{3.6}
\end{equation}
$\Omega _{1}^{2}=\Omega ^{2}+\nu ^{2}.$

Now, from Eq. (3.4b) we can get expression for the solution of Eqs. (3.2),
(3.3):

\begin{equation}
\hat{f}\left( \vec{\varkappa},\vec{k},t\right) =\hat{f}\left( \vec{\varkappa}%
,\vec{k}^{\prime }\left( t^{\prime }=0\right) ,t=0\right) \exp \left\{
-D\int\limits_{0}^{t}dt^{\prime }\left| \vec{k}^{\prime }\left( t^{\prime
}\right) \right| ^{\alpha }\right\} .  \tag{3.7}
\end{equation}
In Eq. (3.7) $\vec{k}^{\prime }\left( t^{\prime }\right) $ is expressed
through $\vec{k}$ with the use of Eqs. (3.5), (3.6). In the next Sections we
consider the peculiarities of the relaxation process and of stationary
states realized in the framework of this solution.

\strut

\section{\strut Homogeneous relaxation and stationary states}

In this Section we consider homogeneous relaxation, $\partial /\partial \vec{%
r}=0$ in Eq. (2.4). Obviously, it corresponds to the particular case $\vec{%
\varkappa}=0$ in the equations of Section 3. Setting $\vec{\varkappa}=0$ in
Eqs. (3.5)-(3.8) we get

\begin{equation}
\hat{f}\left( \vec{k},t\right) =\hat{f}\left( \vec{\varkappa}=0,\vec{k}%
,t\right) =\exp \left\{ i\vec{k}_{0}\vec{v}_{0}-\frac{D}{\alpha \nu }\left(
1-e^{-\alpha \nu t}\right) \left| \vec{k}\right| ^{\alpha }\right\} ,\, 
\tag{4.1}
\end{equation}
where

\begin{equation}
D_{\vec{v}}=\frac D{\alpha \nu }\left( 1-e^{-\alpha \nu t}\right) , 
\tag{4.2}
\end{equation}

\begin{equation}
\vec{k}_{0}=e^{-\nu t}\left\{ \left( \vec{k}\vec{b}\right) \vec{b}+\left( 
\vec{b}\times \left( \vec{k}\times \vec{b}\right) \right) \cos \Omega
t+\left( \vec{b}\times \vec{k}\right) \sin \Omega t\right\} .  \tag{4.3}
\end{equation}

Let us consider the case $\vec{v}_{0}=0$ for the sake of simplicity. The
relaxation process, whose characteristic function is given by Eq. (4.1), is
not an $\alpha $-stable process with independent increments, since $D_{\vec{v%
}\text{ }}$ is not a linear function of time, see Eq. (4.2). The stable
process arises asymptotically at small times only,

\begin{equation}
t<<\tau _{v}=1/\alpha \nu ,  \tag{4.4}
\end{equation}
when the exponent in Eq. (4.2) can be expanded into power series. On the
other hand, after exponential relaxation to the stationary state, that is,
at $t>>\tau _{v}$ the stochastic process $\vec{v}\left( t\right) $ becomes
asymptotically stationary process with the stable PDF independent of $t,$

\begin{equation}
f_{st}\left( \vec{k}\right) =\exp \left( -\frac{D}{\alpha \nu }\left| \vec{k}%
\right| ^{\alpha }\right) .  \tag{4.5}
\end{equation}
We also note that stationary PDF does not depend on the magnetic field.

Another interesting point is related to stationary solutions of fractional
kinetic equations. In the theory of Brownian motion equilibrium Maxwell PDF
over velocity is reached at $t>>1/\nu .$ It is characterized by the
temperature $T$ of surrounding medium. The following relation exists between
the parameter $D$ and the friction coefficient $\nu $ :

\begin{equation}
D=\frac{\nu k_{B}T}{m}\quad ,  \tag{4.6}
\end{equation}
where $k_{B}$ is the Boltzmann constant. The temperature $T$ is a measure of
a mean kinetic energy of the Brownian particle:

\begin{equation}
\left\langle E\right\rangle =\frac{m\left\langle v^{2}\right\rangle }{2}=%
\frac{k_{B}T}{2}\quad .  \tag{4.7}
\end{equation}

\strut Equations (4.6) and (4.7) are examples of fluctuation-dissipation
relations. For this case the source in the Langevin equation is called the
source of internal fluctuations. Relations (4.6) and (4.7) may not be
fulfilled, as it takes place, e.g., in auto-oscillation systems \cite
{Klimontovich}. In such a case one says that there is the source of external
(relatively to the system considered) fluctuations in Eq. (2.2). However,
Maxwell exponential form of stationary solutions retains. As to the Levy
motion, the fluctuation - dissipation relations can not be fulfilled, at
least, because of the infinity of square velocity: $\left\langle
v^{2}\right\rangle =\infty $ for $0<\alpha <2.$ Therefore, we can only say
about the Langevin source as about the source of external fluctuations.
Moreover, the stationary solutions do not possess Maxwell form but instead
more general form of stable distributions. We also note that at present
there is no theory of equilibrium state basing on stable PDFs.

\strut We further study energy distribution in the stationary state,

\begin{eqnarray}
f_{st}\left( E\right) &=&\int d\vec{v}f_{st}\left( \vec{v}\right) \delta
\left( E-\frac{m\vec{v}^2}2\right) =  \tag{4.8} \\
&=&\int d\vec{v}\delta \left( E-\frac{m\vec{v}^2}2\right) \int \frac{d\vec{k}%
}{\left( 2\pi \right) ^d}\exp \left( -i\vec{k}\vec{v}\right) f_{st}\left( 
\vec{k}\right) .  \nonumber
\end{eqnarray}

We recall, that the two possibilities can be considered, namely, (i) the
random electric field is isotropic in the plane perpendicular to the
external magnetic field, and (ii) the field is isotropic in the
three-dimensional space. In the former case $\vec{k}$ and $\vec{v}$ are
two-dimensional vectors in Eq. (4.8), $d=2$, whereas in the latter case $%
\vec{k}$ and $\vec{v}$ are three-dimensional vectors, $d=3.$ We consider
both cases. After some transforms we get from Eq. (4.8),

\begin{equation}
f_{st}\left( E\right) =\frac{1}{m}\int\limits_{0}^{\infty }dk\cdot
kJ_{0}\left( k\sqrt{\frac{2E}{m}}\right) \exp \left( -\mathcal{D}k^{\alpha
}\right) ,\,\,\,\,d=2,  \tag{4.9a}
\end{equation}
\begin{equation}
f_{st}\left( E\right) =\frac{2}{\pi m}\int\limits_{0}^{\infty }dk\cdot k\sin
\left( k\sqrt{\frac{2E}{m}}\right) \exp \left( -\mathcal{D}k^{\alpha
}\right) ,\,\,\,\,d=3,  \tag{4.9b}
\end{equation}
where $\mathcal{D}=D/\alpha \nu .$ The integrals in Eq. (4.9) can be easily
calculated in two particular cases:

1. $\alpha =2$%
\begin{equation}
f_{st}\left( E\right) =\frac{1}{2m\mathcal{D}}\exp \left( -\frac{E}{2m%
\mathcal{D}}\right) ,\,\,\,d=2,  \tag{4.10a}
\end{equation}

\begin{equation}
f_{st}\left( E\right) =\frac{\sqrt{E}}{\sqrt{2\pi }\left( m\mathcal{D}%
\right) ^{3/2}}\exp \left( -\frac{E}{2m\mathcal{D}}\right) ,\,\,d=3, 
\tag{4.10b}
\end{equation}
which are the well-known results of the theory of Brownian motion \cite
{Klimontovich}.

2. $\alpha =1.$

\begin{equation}
f_{st}\left( E\right) =\frac 12\left( \frac{m\mathcal{D}^2}2\right)
^{1/2}\frac 1{\left( E+m\mathcal{D}^2/2\right) ^{3/2}}\,\,,\,\,\,d=2, 
\tag{4.11a}
\end{equation}

\begin{equation}
f_{st}\left( E\right) =\frac{\sqrt{2m\mathcal{D}^{2}}}{\pi }\frac{\sqrt{E}}{%
\left( E+m\mathcal{D}^{2}/2\right) ^{2}}\,\,,\,\,\,\,\,\,d=3.  \tag{4.11b}
\end{equation}
Since the $\alpha $-stable distribution with $\alpha =1$ is called the
Cauchy distribution, Eq. (4.11) corresponds to the case of the Cauchy motion.

From Eq. (4.9) it follows that at large energies $f_{st}\left( E\right) $
has a power law asymptotics for $0<\alpha <2,$

\begin{equation}
f_{st}\left( \vec{k}\right) \varpropto E^{-\left( 1+\alpha /2\right) }\quad ,
\tag{4.12}
\end{equation}
and, thus only the moments of the order $q$ less than $\alpha /2$ are finite
for $\alpha <2$. For the moments of the energy, 
\begin{equation}
\left\langle E^{q}\right\rangle =\stackrel{\infty }{\stackunder{0}{\int }}%
dEE^{q}f_{st}(E)  \tag{4.13a}
\end{equation}
we get 
\begin{equation}
\left\langle E^{q}\right\rangle =\left( 2m\right) ^{q}\mathcal{D}^{2q/\alpha
}\Gamma \left( 1+q\right) \frac{\Gamma \left( 1-2q/\alpha \right) }{\Gamma
\left( 1-q\right) },\,\,\,\,\,d=2,  \tag{4.13b}
\end{equation}
\begin{equation}
\left\langle E^{q}\right\rangle =\frac{2}{\sqrt{\pi }}\left( 2m\right) ^{q}%
\mathcal{D}^{2q/\alpha }\frac{\sin \pi q}{\sin 2\pi q/\alpha }\frac{\Gamma
\left( q\right) \Gamma \left( 3/2+q\right) }{\Gamma \left( 2q/\alpha \right)
^{{}}}\,,\,\,\,\,\,d=3,  \tag{4.13c}
\end{equation}
where $q<\alpha /2<1.$ The particular cases $\alpha =2$ and $\alpha =1$ can
be also obtained from Eqs. (4.13) or by direct using $f_{st}\left( E\right) $
from Eqs. (4.10), (4.11):

$\alpha =2.$%
\begin{equation}
\left\langle E^{q}\right\rangle =\left( 2m\mathcal{D}\right) ^{q}\Gamma
\left( 1+q\right) ,\,\,\,d=2,  \tag{4.14a}
\end{equation}

\begin{equation}
\left\langle E^{q}\right\rangle =\frac{\left( m\mathcal{D}\right) ^{q}2^{q+1}%
}{\sqrt{\pi }}\Gamma \left( q+\frac{3}{2}\right) ,\,\,\,d=3,  \tag{4.14b}
\end{equation}
for all $q\geq 0.$

$\alpha =1.$%
\begin{equation}
<E^{q}>=\left( \frac{m\mathcal{D}^{2}}{2}\right) ^{q}\frac{1}{\sqrt{\pi }}%
\Gamma \left( 1+q\right) \Gamma \left( \frac{1}{2}-q\right)
,\,\,\,\,\,\,\,\,\,d=2,  \tag{4.15a}
\end{equation}

\begin{equation}
<E^{q}>=\frac{\left( m\mathcal{D}^{2}\right) ^{q}}{\pi 2^{q-1}}\Gamma \left( 
\frac{3}{2}+q\right) \Gamma \left( \frac{1}{2}-q\right) ,\,\,\,\,\,\,\,\,d=3,
\tag{4.15b}
\end{equation}
for $q<1/2.$

\strut We carry out numerical simulation based on the solution of the
Langevin equations (2.1) with a two-dimensional isotropic white Levy noise $%
\mathcal{\vec{E}}\left( t\right) $. The case of a strong magnetic field is
simulated. In Fig. 1 typical dependencies $E\left( t\right) $ ($t$ is a
discrete time, $\Delta t=10^{-3}$ is a time step) are shown on the left for
a) $\alpha =1.95$ and c) $\alpha =1.1.$ With Levy index decreasing the
''jumps'' on the trajectories, or ''Levy flights'', become larger; this
effect is due to the power law tails of the PDF of the white Levy noise in
the Langevin equation. In this Figure and below, in Figs. 2 - 4, the
parameters used in simulation are $\Omega =2,\,\,\nu =0.07,$ and $D=1.$ At
the right, Figs. b) and d), the ''Levy flights'' are shown on the $%
(v_{x},v_{y})$ plane. Again, large ''jumps'' are clearly seen in the bottom
figure.

\strut In Fig. 2 the stationary PDFs $f_{st}\left( E\right) $ are shown for
a two-dimensional problem at the left, Figs. a), c) and for a
three-dimensional problem at the right, Figs. b), d), respectively. At the
top, in Fig. a), b), the linear scale is used, whereas in the bottom, in
Figs. c), d), the PDFs are shown in the log-log scale.

The PDFs estimated according to Eqs. (4.9) for the Levy index $\alpha =1.1$
are depicted by solid lines, whereas the PDFs estimated according to Eqs.
(4.10) for $\alpha =2.0$ (Brownian motion) are depicted by dotted lines. The
bottom figures clearly show the power asymptotics of the PDFs. The black
points on the left figures indicate the PDF obtained in numerical simulation
of a two-dimensional problem, the parameters used in numerical simulation
are the same as in Fig. 1. The quantitative agreement between analytical and
numerical results is obvious.

In Fig. 3, as the result of numerical solution of the Langevin equations,
the moments $\left\langle E^{q}\right\rangle $ of the energy are shown for
the Levy index $\alpha =1.6$ and for different orders $q$, see from the
bottom to the top: $q=0.12,$ which is less than $\alpha /2,$ $q=0.8,$ which
is equal $\alpha /2,$ and $q=2.0,$ which is greater than $\alpha /2.$ The
stationary level of the $q$-th moment, estimated according to Eq. (4.13b) is
indicated by dotted line in the bottom figure. It is seen that, with $q$
increasing, the moments of the energy strongly fluctuate; this is the
numerical manifestation of the fact that these moments diverge at $q\geq
\alpha /2.$

\strut

\section{Non-Homogeneous Relaxation and Superdiffusion}

\strut

We turn to the relaxation in non-homogeneous case. Since general analysis of
Eqs. (3.5)-(3.8) is rather complicated and taking in mind that we already
have information about velocity relaxation, we study evolution of a simpler
PDF instead of $f(\vec{r},\vec{v},t)$, namely,

\begin{equation}
f\left( \vec{r},t\left| \vec{r}_{0},\vec{v}_{0}\right. \right) =\int d\vec{v}%
f\left( \vec{r},\vec{v},t\left| \vec{r}_{0},\vec{v}_{0}\right. \right) , 
\tag{5.1}
\end{equation}
whose characteristic function is

\begin{equation}
\hat{f}\left( \vec{\varkappa},t\left| \vec{r}_{0},\vec{v}_{0}\right. \right)
=\hat{f}\left( \vec{\varkappa},\vec{k}=0,t\left| \vec{r}_{0},\vec{v}%
_{0}\right. \right) ,  \tag{5.2}
\end{equation}
and the characteristic function in the r.h.s. of Eq. (5.2) is given by Eqs.
(3.5)-(3.8). Putting $\vec{k}=0$ in Eqs. (3.5), we get 
\begin{eqnarray}
\vec{k}^{\prime }\left( t^{\prime }\right) &=&\frac{\left( \vec{\varkappa}%
\vec{b}\right) \vec{b}}{\nu }\left[ 1-e^{\nu \left( t^{\prime }-t\right)
}\right] +  \nonumber \\
&&+\frac{\nu }{\Omega _{1}^{2}}\left( \vec{b}\times \left( \vec{\varkappa}%
\times \vec{b}\right) \right) \left\{ 1-e^{\nu (t^{\prime }-t)}\left[ \cos
\Omega \left( t^{\prime }-t\right) +\frac{\Omega }{\nu }\sin \Omega \left(
t^{\prime }-t\right) \right] \right\} +  \nonumber \\
&&+\frac{\Omega }{\Omega _{1}^{2}}\left( \vec{b}\times \vec{\varkappa}%
\right) \left\{ 1-e^{\nu (t^{\prime }-t)}\left[ \cos \Omega \left( t^{\prime
}-t\right) -\frac{\nu }{\Omega }\sin \Omega \left( t^{\prime }-t\right)
\right] \right\} .  \tag{5.3}
\end{eqnarray}

In the absence of the magnetic field, $\vec{B}=0\,,$ we get, using Eqs.
(5.3), (5.2)and (3.7),

\begin{equation}
\hat{f}\left( \vec{\varkappa},t\left| \vec{r}_{0},\vec{v}_{0}\right. \right)
=\exp \left\{ i\vec{\varkappa}\vec{r}_{0}+\frac{i\vec{\varkappa}\vec{v}_{0}}{%
\nu }\left( 1-e^{-\nu t}\right) -\frac{D\left| \vec{\varkappa}\right|
^{\alpha }}{\nu ^{\alpha }}\int\limits_{0}^{t}d\tau \left( 1-e^{-\nu
t}\right) ^{\alpha }\right\} \,.  \tag{5.4}
\end{equation}
For one-dimensional case this result was obtained in \cite{Chechkin3}.

For the case of a strong magnetic field, $\Omega >>\nu $ , we get, again by
using Eqs. (5.3), (5.2) and (3.7),

\begin{equation}
\hat{f}\left( \vec{\varkappa},t\left| \vec{r}_{0},\vec{v}_{0}\right. \right)
=\exp \left\{ i\vec{\varkappa}\vec{r}_{0}+i\vec{k}_{0}\vec{v}%
_{0}-D\int\limits_{0}^{t}dt^{\prime }\left| \vec{k}^{\prime }\left(
t^{\prime }\right) \right| ^{\alpha }\right\} \,\,,  \tag{5.5}
\end{equation}
where

\begin{eqnarray}
\vec{k}^{\prime }\left( t^{\prime }\right) &=&\frac{\left( \vec{\varkappa}%
\vec{b}\right) \vec{b}}{\nu }\left( 1-e^{-\nu \tau }\right) +\frac{e^{-\nu
\tau }}{\Omega }\left( \vec{b}\times \left( \vec{\varkappa}\times \vec{b}%
\right) \right) \sin \Omega \tau +  \tag{5.6} \\
&&+\frac{1-e^{-\nu \tau }\cos \Omega \tau }{\Omega }\left( \vec{b}\times 
\vec{\varkappa}\right) ,\,\,\,\tau =t-t^{\prime },  \nonumber
\end{eqnarray}
and $\vec{k}_{0}$ is given by Eq. (5.6) at $t^{\prime }=0$ $\left( \text{%
that is, }\tau =t\right) \,.$

For simplicity we put $\vec{r}_{0}=0$ , and $\vec{v}_{0}=0$ , as in previous
Section. Further, we are interested in diffusion across the magnetic field
and, therefore we set $\left( \vec{\varkappa}\vec{b}\right) =0$ in Eq.
(5.6), thus not considering the motion of a particle along the magnetic
field. For the characteristic function we get

\begin{equation}
\hat{f}\left( \vec{\varkappa},t\right) \equiv \hat{f}\left( \vec{\varkappa}%
,t\left| 0,0\right. \right) =\exp \left\{ -\frac{D}{\Omega ^{\alpha }}%
\varkappa ^{\alpha }\int\limits_{0}^{t}d\tau \left( 1-2e^{-\nu \tau }\cos
\Omega \tau +e^{-2\nu \tau }\right) ^{\alpha /2}\right\} \,\,,  \tag{5.7}
\end{equation}
where $\vec{\varkappa}$ is two-dimensional vector in the plane perpendicular
to $\vec{B}$ , $\varkappa \equiv \left| \vec{\varkappa}\right| $ .

At $\tau >>1/\nu $ (the diffusion stage of relaxation) the expression (5.7)
gets a simple form,

\begin{equation}
\hat{f}\left( \vec{\varkappa},t\right) =\exp \left\{ -\frac{D}{\Omega
^{\alpha }}\varkappa ^{\alpha }t\right\} \,\,,  \tag{5.8}
\end{equation}
which is the characteristic function of an $\alpha $-stable isotropic
process, compare with Eq. (2.3). Now we consider the PDF and diffusion
process in more detail. By taking inverse Fourier transform from Eq. (5.8)
we get the PDF,

\begin{equation}
f\left( \vec{r},t\right) =\int \int \frac{d\vec{\varkappa}}{4\pi ^2}e^{-i%
\vec{\varkappa}\vec{r}}\hat{f}\left( \vec{\varkappa},t\right)
=\int\limits_0^t\frac{d\varkappa \cdot \varkappa }{2\pi }J_0\left( \varkappa
r\right) \exp \left( -\frac{Dt}{\Omega ^\alpha }\varkappa ^\alpha \right)
\,\,,\,\,\,\,\,\,r\equiv \left| \vec{r}\right| .  \tag{5.9}
\end{equation}

The particular cases of Eq. (5.9) are

1. $\alpha =2,$

\begin{equation}
f\left( \vec{r},t\right) =\frac{\Omega ^{2}}{4\pi Dt}\exp \left( -\frac{%
\Omega ^{2}}{4Dt}r^{2}\right) \,,  \tag{5.10}
\end{equation}
and

2. $\alpha =1$,

\begin{equation}
f\left( \vec{r},t\right) =\frac{Dt/\Omega }{2\pi \left( r^{2}+\left(
Dt/\Omega \right) ^{2}\right) ^{3/2}}\quad \,.  \tag{5.11}
\end{equation}
In general case $0<\alpha <2$ the asymptotics of the PDF at large $r$
behaves as

\begin{equation}
f\left( \vec{r},t\right) \varpropto \frac{Dt}{\Omega ^{\alpha }r^{2+\alpha }}%
\,\quad .  \tag{5.12}
\end{equation}
\vspace{0in}\newpage

It implies that the $q$-th moments of $r$ diverge at $q\geq \alpha $ . The
expression for the moments follows from Eq. (5.9): 
\begin{equation}
\left\langle r^{q}\right\rangle \equiv M_{r}\left( t;q,\alpha \right)
=\left[ \frac{\left( Dt\right) ^{1/\alpha }}{\Omega }\right] ^{q}C_{2}\left(
q;\alpha \right) \quad ,  \tag{5.13}
\end{equation}
where 
\begin{equation}
C_{2}\left( q;\alpha \right) =\int\limits_{0}^{\infty }d\varkappa
_{1}\varkappa _{1}^{1+q}\int\limits_{0}^{\infty }d\varkappa _{2}\varkappa
_{2}J_{0}\left( \varkappa _{1}\varkappa _{2}\right) e^{-\varkappa
_{2}^{\alpha }}\quad .  \tag{5.14}
\end{equation}
The integral over $\varkappa _{2}$ behaves as $\varkappa _{1}^{-2-\alpha }$
at large $\varkappa _{1}$ , thus $C_{2}\left( q;\alpha \right) $ diverges at
upper limit of $\varkappa _{1}$ at $q\geq \alpha .$ The particular cases
following from Eqs. (5.13), (5.14) are as follows:

1. $\alpha =2,$

\begin{equation}
\left\langle r^{q}\right\rangle =\left( \frac{4D}{\Omega ^{2}}\right)
^{q/2}t^{q/2}\Gamma (1+\frac{q}{2}),\,\,\,\,\,q>0\quad ,  \tag{5.15}
\end{equation}
and

2. $\alpha =1$ ,

\begin{equation}
\left\langle r^{q}\right\rangle =\frac{1}{2}\left( \frac{Dt}{\Omega }\right)
^{q}\Bbb{B}\left( 1+\frac{q}{2};\frac{1}{2}-\frac{q}{2}\right)
,\,\,\,\,\,0<q<1  \tag{5.16}
\end{equation}
where $\Bbb{B}$ is the beta-function.

From Eq.(5.15) we get a classical diffusion law for the square displacement
of charged particle across the magnetic field: 
\begin{equation}
\left\langle r^{2}\right\rangle \varpropto \frac{t}{B^{2}}  \tag{5.17}
\end{equation}
One can introduce ''the typical displacement'' of a particle defined as 
\begin{equation}
\Delta r=\left\langle r^{q}\right\rangle ^{1/q}\quad .  \tag{5.18}
\end{equation}
From Eq. (5.15) it follows that at any $q$ (not only at $q=2$) we have for
the Brownian particle,

\begin{equation}
\Delta r\varpropto \frac{t^{1/2}}{B}\quad ,  \tag{5.19}
\end{equation}
with the prefactor which, of course, depends on $q$. We recall that usually
just $t$ and (especially in plasma physics) $B$-dependences serve as
indicator of normal or anomalous diffusion. Therefore, we can use the
typical displacement (5.18) as a measure of anomalous diffusion rate at $%
0<\alpha <2$ and any $q<\alpha $. Indeed, it follows from Eq. (5.13) that
for the anomalous diffusion

\begin{equation}
\Delta r\equiv \left\langle r^{q}\right\rangle ^{1/q}\varpropto \frac{%
t^{1/\alpha }}{B}\,\,\,.  \tag{5.20}
\end{equation}
Expression (5.20) teaches us that in our model diffusion is anomalously fast
over $t$ , since $\alpha <2$ ; this diffusion is also called superdiffusion.
At the same time it retains classical scaling over $B$ ( that is, given by
Eqs. (5.17) or (5.19)). We also note that the obtained $t$ - dependence is
the typical superdiffusion law within the framework of fractional kinetics,
see Refs. \cite{Jespersen}, \cite{Uchaikin}, \cite{Chechkin3}, \cite
{Chechkin4}.

\strut Basing on the numerical solution to the Langevin equations (2.1) we
estimated numerically the moments $M_{r}\left( t;q,\alpha \right) $ by
averaging over 100 realizations, each consisting from 50000 time steps. In
Fig. 4 the $q$-th root of the $q$-th moment (characteristic displacement $%
\Delta r$) is shown versus $t$ in a log-log scale for the three values of $q$%
, and for the Levy index equal $1.2$, see from bottom to top: $q=0.12,\,$%
which is less than $\alpha ,$ $q=1.1$ which is nearly $\alpha $, and $q=2.0$
(variance), which is greater than $\alpha .$ The dashed lines have the slope 
$1/\alpha ,$ which is, in fact, the theoretical value of the diffusion
exponent at $q<\alpha ,$ see Eq. (5.20). At $q<\alpha $ the numerical curve
is well fitted by the dotted line. At $q\geq \alpha $ theoretical value of
the moment is infinite, and in numerical simulation the moment strongly
fluctuates.

In our numerical simulation the scaling given by Eq. (5.20) was also checked
in more detail. The results are presented in Fig. 5. At the left, in Figure
a), we show the $1/B$ - dependence of the characteristic displacement $%
\Delta r.$ The Levy index $\alpha $ is equal $1.2,$ and $q$ is equal 0.12.
The values of $\Delta r$ obtained in numerical simulation are shown by black
points, which are well fitted by straight dotted line. This fact confirms
that $\Delta r$ is inversely proportional to the magnetic field. At the
right in Figure b) we show the exponent $\mu $ in the relation $\Delta r\sim
t^{\mu }$ versus $1/\alpha .$ The black points indicate the result of
numerical simulation, whereas the dotted line shows the straight line $\mu
=1/\alpha $. We conclude that the right figure confirms the obtained
theoretical dependence $\Delta r\sim t^{1/\alpha }.$

\strut

\section{\strut Conclusion}

\strut \ In this paper we propose fractional Fokker-Planck equation (FFPE)
for the description of the motion of a charged particle in constant magnetic
field and stochastic electric fields. The latter is assumed to be a white
Levy noise. We also assume that the particle is influenced by a linear
friction force. Such formulation is a natural generalization of the
classical problem for the Brownian particle \cite{Korsunoglu}. It allows us
to consider in a simplest way the peculiarities of the motion stipulated by
non-Gaussian Levy statistics of a random electric field.

The main results are as follows:

1. The general solution to FFPE for a charged particle in constant magnetic
field is obtained. In case of the absence of magnetic field this solution
lead to the results obtained previously. However, the general solution also
allows us to study the opposite case of a strong magnetic field in detail.

2. The properties of stationary states are studied for two- and
three-dimensional motions. The velocity relaxation is studied in detail, and
non-Maxwellian stationary states are found, for which the velocity PDF is a
Levy stable distribution.

The energy PDFs are obtained, which have power law tails. This circumstance
leads to divergence of the energy mean. In the real experiments as well as
in numerical simulation the divergence manifests itself in large
fluctuations of the mean value with time.

3. The superdiffusion of a charged particle across magnetic field is studied
within the framework of FFPE. The fractional moments of space displacement
are estimated, and anomalous dependence of characteristic displacement $%
\Delta r$ versus time, $\Delta r\sim t^{1/\alpha },$ $\alpha <2\,,$ is
found. The typical displacement is inversely proportional to the magnetic
field. Therefore the diffusion described by our FFPE demonstrate anomalous
behavior with time and remains classical with respect to the magnetic field
dependence.

4. We carry out numerical simulation based on the solution to the Langevin
equations for a charged particle in constant magnetic and random electric
fields. We study numerically the process of relaxation and stationary energy
states for different Levy indexes as well as superdiffusion process. The
results of numerical simulation are in qualitative agreement with analytical
estimates.

\strut In summary, we believe that fractional kinetics will be a useful
complementary tool for understanding and description of variety of
non-equilibrium phenomena in space and thermonuclear plasmas. More
elaborated fractional kinetic equations based on more sophisticated Langevin
equations can be constructed, which, in particular, lead to a finite
variance of the displacement of a particle and to anomalous $B$ -
dependence. On the other hand, a consistent development of the theory of
essentially non-Gaussian plasma fluctuations is also of interest.

\strut

\strut

{\Large Acknowledgments}

\strut

This work is financed by INTAS Call 2000, Project No. 847. One of us (M.S.)
was supported by the KBN grant 2P03D01417.

\strut

\strut

\strut \bigskip

\strut

\strut

\strut

\strut

\strut

{\large Figure Captions}

Fig. 1. Numerical solution to the Langevin equations (2.1). At the left:
typical dependences of the energy $E$ versus discrete time $t$ for a) $%
\alpha =2.0$ (Gaussian noise term in the Langevin equations), and c) $\alpha
=1.1.$ At the right: ($v_{x}-v_{y}$) plane for b) $\alpha =2.0$, and d) $%
\alpha =1.1.$

Fig. 2. Stationary energy PDFs are shown in linear scale (at the top) and in
log-log scale (in the bottom), for two-dimensional motion (in the left) and
three-dimensional motion (in the right), respectively. The PDF with $\alpha
=1.1$ is shown by solid curves, the PDF with $\alpha =2$ is shown by dotted
lines. The points indicate the PDF obtained in numerical simulation.

Fig. 3. The results of numerical simulation based on the solution to the
Langevin equations. The $q-$th moments of the energy vs $t$ for different
orders of $q$ and for $\alpha =1.6.$ From bottom to top: $q=0.12<\alpha /2;\
q=0.8=\alpha /2;$ $q=2.0>\alpha /2.$ In the bottom figure the dotted
vertical line indicates the velocity relaxation time $\tau _{v}$, whereas
the dotted horizontal line shows the analytical value of the moment.

Fig. 4. The results of numerical simulation based on the solution to the
Langevin equations. The $q$-th root of the $q$-th moment as a function of $t$
in a log-log scale for different orders of $q$ and for $\alpha =1.2.$ From
bottom to top: $q=0.12<\alpha ;q=1.2=\alpha ;q=2.0>\alpha .$ The tangent of
a slope of dashed lines equals to $1/\alpha .$

Fig. 5. At the left: the $q$-th root of the $q$-th moment versus $1/B.$ The
black points, which are fitted by straight dotted line, show the result of
numerical simulation. At the right: the power $\mu $ in the relation $%
\left\langle r^{q}\right\rangle ^{1/q}\propto t^{\mu }$ versus $\alpha .$
The black points, which are fitted by straight dotted line result from
numerical simulation.

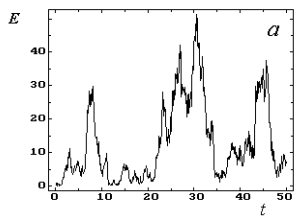
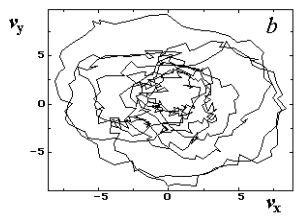
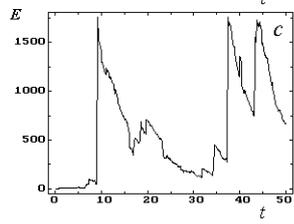
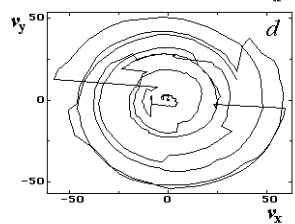
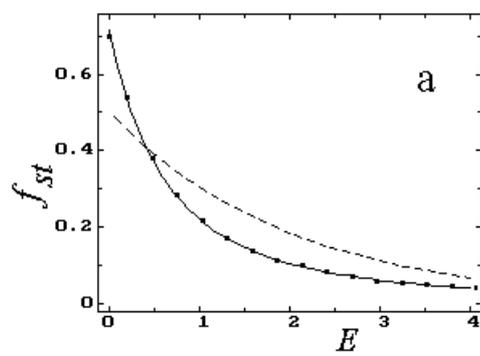
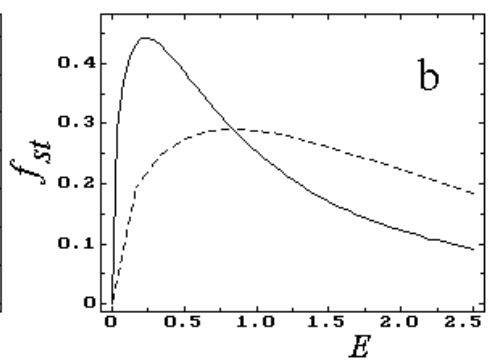
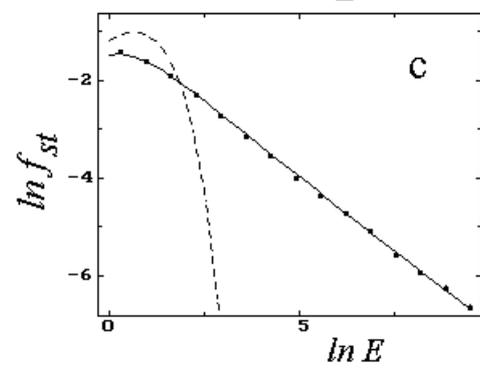
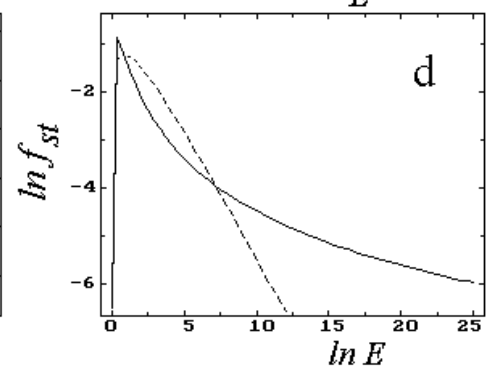
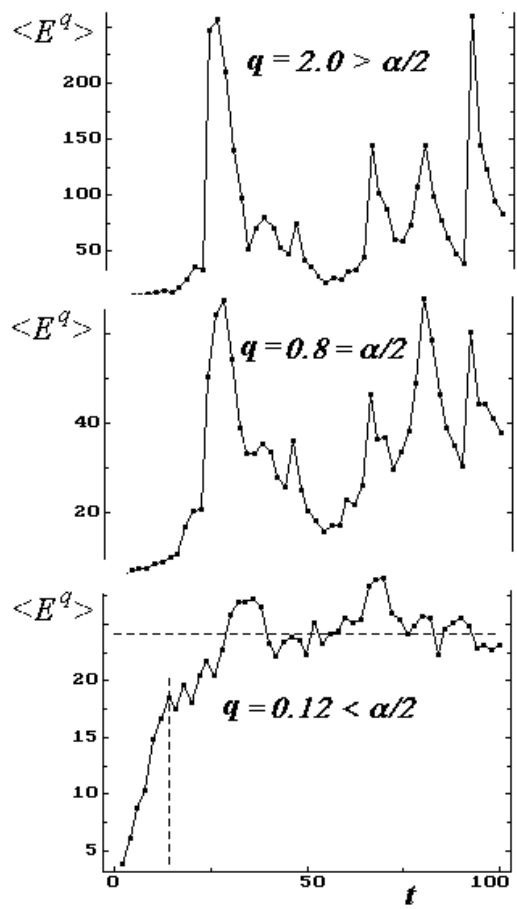
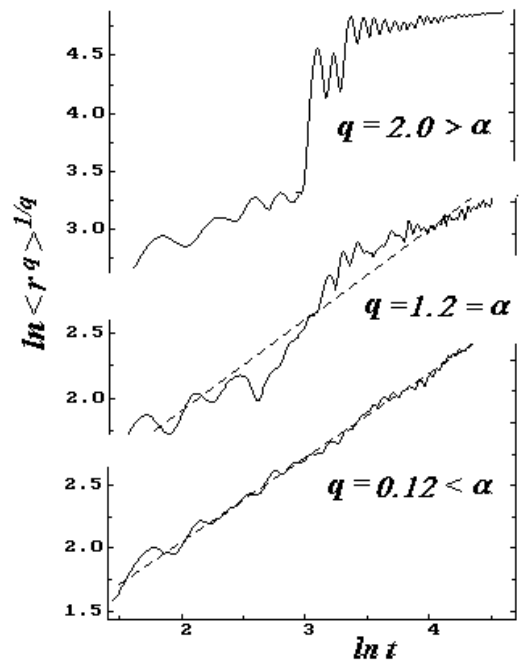
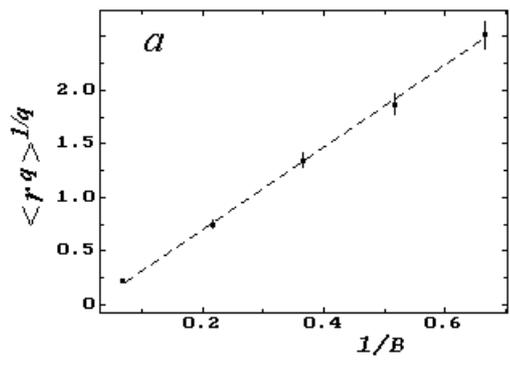
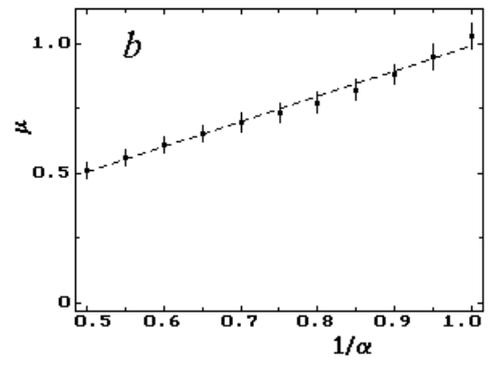
\end{document}